\documentclass[letterpaper,twocolumn,showpacs,preprintnumbers,amsmath,amssymb]{revtex4}

\usepackage{graphicx}
\usepackage{dcolumn}
\usepackage{bm}

\begin{document}

\preprint{Accepted for publication in {\em Phys. Rev. Lett.}.
Titles of references and e-print numbers added.}

\title{Spectroscopic observation of resonant electric dipole-dipole
interactions between cold Rydberg atoms}%

\author{K.~Afrousheh}
\author{P.~Bohlouli Z.}
\author{D.~Vagale}
\author{A.~Mugford}
\author{M.~Fedorov}
\author{J.~D.~D.~Martin}
\affiliation{%
Department of Physics and Institute for Quantum Computing \\
University of Waterloo, Waterloo, ON, N2L 3G1, Canada
}%

\date{November 8th, 2004}

\begin{abstract}
Resonant electric dipole-dipole interactions between cold Rydberg atoms
were observed using microwave spectroscopy.
Laser-cooled $^{85}$Rb atoms in a magneto-optical trap were optically
excited to $45d_{5/2}$ Rydberg states using a pulsed laser.  
A microwave pulse transferred
a fraction of these Rydberg atoms to the $46p_{3/2}$ state.
A second microwave pulse then drove atoms in the 
$45d_{5/2}$ state to the $46d_{5/2}$ state, and was used
as a probe of interatomic interactions.
The spectral
width of this two-photon probe transition was found to depend on the 
presence of the $46p_{3/2}$ atoms, and is due to the resonant
electric dipole-dipole interaction between $45d_{5/2}$ and 
$46p_{3/2}$ Rydberg atoms.
\end{abstract}


\maketitle

The vast separation of the electron and ion-core in high-$n$ 
Rydberg atoms is responsible for their large transition dipole moments
\cite{gallagher:1994}.  
These dipole moments dictate the strength
of the dipole-dipole interaction between pairs of atoms.
Therefore, excitation to Rydberg states
allows one to turn on strong interactions between atoms which
would otherwise be negligible.  This has recently received 
considerable attention in the context of quantum information
processing with cold neutral atoms 
\cite{jaksch:2000,lukin:2001,
protsenko:2002,saffman:2002,safronova:2003,ryabtsev:2004}.
For example, it has been proposed that a single
excited Rydberg atom in a cloud may block further resonant
excitation due to the dipole-dipole
interaction -- a process known as ``dipole blockade'' \cite{lukin:2001}.
This would allow clouds of cold atoms to 
store qubits without the addressing of individual 
atoms and may also be useful for
constructing single-atom and single-photon sources \cite{saffman:2002}.

Non-resonant dipole-dipole (van der Waals) interactions 
between Rydberg atoms were first observed by 
Raimond {\it et al.}~\cite{raimond:1981} using spectral line
broadening.
Recently it has been shown that Rydberg excitation densities 
in a magneto-optical trap (MOT) are limited by these interactions
\cite{tong:2004,singer:2004}.
Dipole-dipole interactions
between Rydberg atoms have also been 
studied in the context of resonant energy
transfer \cite{gallagher:1994}.
Of particular
relevance to this work is the observation of resonant energy transfer
between cold Rydberg atoms \cite{anderson:1998,mourachko:1998},
where the
Rydberg atoms behave more like an amorphous solid than a gas,
and one cannot solely consider binary interactions to
explain the transfer process 
\cite{anderson:2002,mourachko:2004}.
However, use of the resonant
dipole-dipole interaction between cold Rydberg atoms 
to influence radiative transitions -- as presented
in this work -- is an unexplored area.

We excite Rydberg states using a pulsed laser
with no stringent demands on linewidth or stability.
Dipole-dipole interactions are then introduced and probed
using microwave transitions between Rydberg states.
This is advantageous since commercial microwave synthesizers 
are readily tunable, highly stable, and have easily
adjustable powers and pulse-widths, as compared
to lasers.  Using this approach we have made the first spectroscopic
observation of the {\em resonant} dipole-dipole interaction between
cold Rydberg atoms using radiative transitions.

To observe interactions between atoms
that are effectively stationary,
we excite Rydberg states of
laser-cooled atoms.  A standard MOT is used 
as a source of cold $^{85}$Rb atoms.
The cooling and trapping light remains on during the experiment
and thus a fraction of atoms are in the excited $5p_{3/2}$ state.
A Littman-Metcalf, Nd:YAG pumped, nanosecond pulsed dye laser is 
used to excite these translationally 
cold $5p_{3/2}$ atoms to the $45d_{5/2}$ state.
The total number of Rydberg atoms excited is
sensitive to the frequency spectrum of the multimode
dye laser -- which varies from shot to shot.
Since we are interested in density dependent effects,
data is recorded for every laser shot and 
processed to select laser shots 
corresponding to specified Rydberg atom densities
\cite{robinson:2000}.  
The fluctuations 
may even be considered advantageous, as they allow a 
range of Rydberg densities to be sampled automatically.

The MOT is formed between two
metal plates 3.6 cm apart.  These plates
contain small holes to let the cooling and trapping lasers through.
During laser excitation, an electric field of 5.1 V/cm is
applied using these plates, which removes ions formed during
laser excitation.  This field is switched off with a 
0.2 $\mu$s fall time approximately 0.1 $\mu$s 
after photoexcitation, and remains less than 0.1 V/cm during
application of the microwave pulses
(this is verified by observing the Stark shifts of
the microwave transitions \cite{osterwalder:1999}).

\begin{figure}
\includegraphics{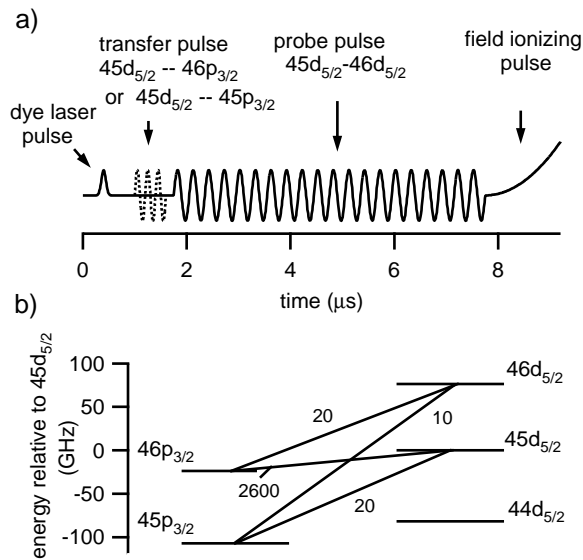}
\caption{ \label{fg:timing}
a) Timing for experiment and b) 
energy levels of relevant states.
The spectroscopic data is from Ref.~\cite{li:2003}.  
The matrix elements magnitudes
$|\!\!<\!\!n\ell_{j}|r|n'\ell'_{j'}\!\!>\!\!|$
are shown between states (in atomic units).
}
\end{figure}

After optical excitation to the 45$d_{5/2}$ state, 
a fraction of the atoms may be transferred to either
the $46p_{3/2}$ state or the $45p_{3/2}$ state using a microwave
pulse to drive a one-photon transition (the ``transfer pulse'')
(see Fig.~\ref{fg:timing}).
The microwave radiation
is introduced into the experimental region by horns placed
outside the vacuum chamber, directed towards the trapped atom cloud 
through a large fused silica viewport.
To obtain 50 \% ($\pm$ 10 \%) transfer of  
Rydberg atoms from the 45$d_{5/2}$ to the $46p_{3/2}$ state
in 0.6 $\mu$s long pulses requires less than 1 mW.

To observe dipole-dipole interactions from the 
line broadening of spectroscopic transitions,
it is desirable to minimize other sources of broadening.
In particular, the inhomogeneous magnetic fields necessary 
for operation of the MOT may broaden spectroscopic transitions 
due to the Zeeman effect.  However, as Li {\it et al.}~\cite{li:2003}
have demonstrated, two-photon transitions
between Rydberg states with the same $g_{J}$ factors 
(eg. $nd_{5/2}-(n+1)d_{5/2}$) show negligible broadening in a MOT. 
Thus, we use  the 45$d_{5/2}-46d_{5/2}$ two-photon
transition as a high resolution, sensitive ``probe'' of interatomic
interactions.  This 6 $\mu$s probe pulse 
requires a total power of less than 100 $\mu$W
and is introduced to the atoms in the same manner 
as the transfer pulses.  
The probe frequency is typically scanned between
laser shots and the Rydberg state populations are measured after 
each shot using the selective field ionization (SFI) technique
\cite{gallagher:1994}.
Absolute, spatially averaged Rydberg densities are obtained
from knife-edge measurements of the dye laser beam waist,
$5p_{3/2}$ fluorescence imaging, and calibration
of the microchannel plate detector.  These estimated densities
are systematically uncertain by a factor of 2.

\begin{figure}
\includegraphics{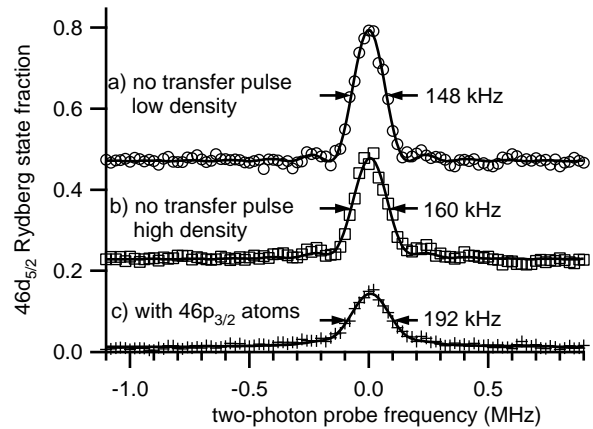}
\caption{\label{fg:spectra}
Observation of the two-photon $45d_{5/2}-46d_{5/2}$ microwave
transition with 
a) ($\circ$)
all Rydberg atoms initially in the $45d_{5/2}$ state
at a density of $2 \times 10^{6}$ cm$^{-3}$ 
b) ($\Box$)
all Rydberg atoms initially in the $45d_{5/2}$ state
at a density of $1 \times 10^{7}$ cm$^{-3}$  
c) ($+$)
   one-half of the Rydberg atoms initially in the $45d_{5/2}$ state and
   the other half in the $46p_{3/2}$ state
   with a total density of $1 \times 10^{7}$ cm$^{-3}$. 
Also shown with solid lines are a) the ${\rm sinc}^2(\pi f T)$ 
lineshape ($T=6 \: \mu s$),
and
b) and c) the same lineshape convolved with a Lorentzian 
of variable width to give the best least-squares fits
(see text).
Spectra are offset vertically for clarity.
The probe frequency
shown on the horizontal axis is twice the applied frequency,
offset by 76.4543 GHz.
}
\end{figure}

Figure \ref{fg:spectra}a) 
shows a microwave spectrum of the two-photon probe transition
without the application of a transfer pulse.
This is well-matched by a superimposed ${\rm sinc}^2(\pi f T)$ 
lineshape suitable for a square excitation pulse of
duration $T= 6 \: \mu s$.
As expected, inhomogeneous Zeeman broadening
makes a negligible contribution to the linewidth \cite{li:2003}.

We enhance the interactions between Rydberg atoms by introducing
a microwave pulse shortly after
photoexcitation, 
transferring 50\% of atoms to the
46$p_{3/2}$ state, before application of the two-photon 
$45d_{5/2}-46d_{5/2}$ probe pulse.
Unlike the two-photon probe pulse, the one-photon transition
is broadened by several MHz due to the inhomogeneous magnetic
field \cite{li:2003}.   We do not observe any Rabi flopping 
for this ``transfer'' transition and consequently do not
investigate the possibility of preparing coherent superpositions of
these two states.  
With a total Rydberg density of $10^7 \: {\rm cm^{-3}}$, 
converting half of the initially excited $45d_{5/2}$ 
atoms to the $46p_{3/2}$ state consistently broadens the linewidth 
of the two-photon $45d_{5/2}-46d_{5/2}$ probe transition 
from $160 \pm 5 \; {\rm kHz}$ to 
$192 \pm 5 \; {\rm kHz}$ -- see Fig.~\ref{fg:spectra}
(all widths in this paper are full-width half-maxima).  

\begin{figure}
\includegraphics{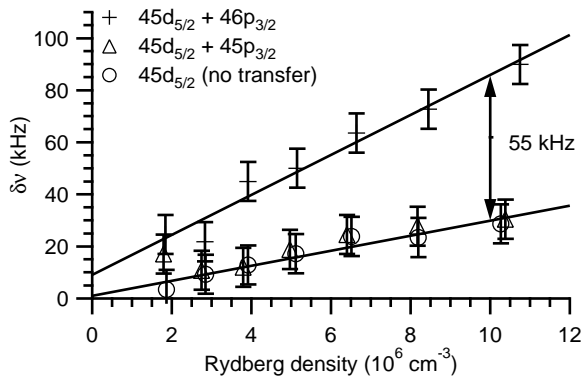}
\caption{\label{fg:ldens}
Broadening of the $45d_{5/2}-46d_{5/2}$ probe transition
as a function of average Rydberg density with and
without the transfer pulses (see text for definition of
$\delta\nu$).
}
\end{figure}

As mentioned previously, fluctuations
in Rydberg state excitation efficiency may be exploited to accumulate
data over a broad range of density conditions.  
To analyze linewidths quantitatively
a ${\rm sinc}^2(\pi f T)$ lineshape ($T=6 \: \mu s$) is
convolved by a Lorentzian with a variable width
$\delta \nu$ adjusted for the best least-squares fit to individual
spectra.  The Lorentzian form is supported by a theoretical
model ({\it vide infra}).
Figure \ref{fg:ldens} shows $\delta \nu$ both
with and without introduction of the 46$p_{3/2}$ atoms, as a
function of average Rydberg density.
With decreasing density $\delta \nu$ approaches zero, suggesting
the influence of interatomic interactions
-- which inevitably weaken at the large average
separations corresponding to low densities.

The difference in linewidths with and without the $46p_{3/2}$
atoms can be attributed to resonant electric dipole-dipole interactions.  
To obtain estimates of the line broadening we
consider interactions between pairs of atoms (A and B) 
due to the electric dipole-dipole interaction operator:
\begin{equation}
\hat{V}_{dd} = \frac{\vec{\mu}_{\rm A} \cdot \vec{\mu}_{\rm B}
    - 3 (\vec{\mu}_{\rm A} \cdot \vec{n}) 
(\vec{\mu}_{\rm B} \cdot \vec{n})}{R_{\rm AB}^3}
\end{equation}
where $\vec{\mu}_{\rm A}$ and 
$\vec{\mu}_{\rm B}$ are the electric dipole matrix element operators evaluated
on each atom, 
$\vec{n}$ is the unit vector pointing between the atoms,
and $R_{\rm AB}$ is the separation of the two atoms.
This perturbation may split the otherwise
energy degenerate states
$|1>=\!\!|45d_{5/2}m_{j,{\rm A1}}\!\!>_{\rm A}\!\!
|46p_{3/2}m_{j,{\rm B1}}\!\!>_{\rm B}$ 
and
$|2>=\!\!|46p_{3/2}m_{j,{\rm A2}}\!\!>_{\rm A}\!\!
|45d_{5/2}m_{j,{\rm B2}}\!\!>_{\rm B}$,
and with a 50\% mixture of $45d_{5/2}$ and $46p_{3/2}$ atoms
we can obtain a very rough idea of the 
magnitude of the energy splittings from 
$\Delta\nu_{dd} \approx \mu^2 / R^3$, where
$\mu = |\!\!<\!\!45d_{5/2,1/2}|\mu_z|46p_{3/2,1/2}\!\!>\!\!|
\approx 0.49 |\!\!<\!\!45d_{5/2}|r|46p_{3/2}\!\!>\!\!|$,
$R=(4\pi n_{46p}/3)^{-1/3}$,
and $n_{46p}$ is the $46p_{3/2}$ number density.
The radial matrix element is evaluated by numerical integration
of the Rydberg electron wave functions 
(see  Fig.~\ref{fg:timing}b) \cite{zimmerman:1979}. 
The 45$d_{5/2}$ 
and 46$p_{3/2}$ states are strongly dipole coupled, whereas the
46$d_{5/2}$ and 46$p_{3/2}$ states are not.  
Therefore, only the initial state of the two-photon
probe transition is split by the 
resonant dipole-dipole interaction with 46$p_{3/2}$ atoms.
A density of $n_{\rm 46p}=5\times 10^6 \: {\rm cm}^{-3}$
gives $\Delta\nu_{dd}=33 \: {\rm KHz}$ -- the same order
of magnitude as the observed broadening (see Fig.~\ref{fg:ldens}).   

The dipole-coupling between
the $45d_{5/2}$ and $45p_{3/2}$ states is much smaller than that 
between the $45d_{5/2}$ and $46p_{3/2}$ states 
(see Fig.~\ref{fg:timing}b)).
This suggests the following test.
Instead of transferring 50\% of the atoms to the $46p_{3/2}$ 
state,
we transfer 50\% of atoms to the $45p_{3/2}$ 
state, and study the broadening
of the probe transition with increasing Rydberg atom density.  Based
on the much smaller dipole matrix element, it is expected that
introducing the $45p_{3/2}$ 
atoms will have little influence on
the linewidth of the probe transition.  
Figure \ref{fg:ldens} shows that the introduction of the $45p_{3/2}$ 
atoms gives linewidths
which are experimentally indistinguishable from the 100\% $45d_{5/2}$
case.

Now we consider a calculation of the linewidths which
accounts for the orientations of the dipole and $\vec{n}$
operators and the distribution in interacting atom separations
-- which were neglected in the simple estimate presented
above.
With no magnetic field, there is a large energy degeneracy
corresponding to the different possible magnetic sub-levels
for the two atoms (A and B) (without $\hat{V}_{dd}$).
However, in the MOT the average Zeeman shifts are relatively
large ($\approx$1 MHz) compared to the influence of $\hat{V}_{dd}$
($<$100 kHz).  Therefore only those 
states that are exactly degenerate 
are strongly coupled  
($m_{j,{\rm A}1}=m_{j,{\rm B}2}$ and 
$m_{j,{\rm B}1}=m_{j,{\rm A}2}$).
In the absence of detailed information
about the magnetic sub-level populations, two extremes are considered:
a) atoms are randomly distributed over all possible magnetic
sub-levels, and
b) all atoms are in $m_j=1/2$.
In the first case the effective interaction is diminished, 
since there will be pairs of atoms which
will not interact at all 
(eg. $m_{j,{\rm A}1}=5/2$, $m_{j,{\rm B}1}=-3/2$).
To simulate the lineshape -- in particular the randomness
associated with $R_{\rm AB}$ and $\vec{n}$ --
we consider a $45d_{5/2}$ atom at
the center of a sphere containing a number
of randomly placed $46p_{3/2}$ atoms.  The matrix elements 
$<1|\hat{V}_{dd}|2>$
are computed using the $45d_{5/2}$ atom and each $46p_{3/2}$
within the sphere, using the magnetic sub-level distribution 
scenarios discussed above.
The matrix element with the largest magnitude is selected $V_{\rm max}$ 
(a binary ``strongest
interacting'' neighbor approximation).  This 
splits the $45d_{5/2}-46d_{5/2}$ transition into a
doublet $\pm V_{\rm max}$ (the energy eigenvalues of our
simplified two-state system).
This process was repeated numerous times,
and the resulting splittings histogramed to obtain
lineshapes, which converged as the size of the
sphere and number of particles increased (we maintained a constant
average density).   These have sharp dips in
their centers, but Lorentzian wings 
(the dips are much sharper than the transform-limited 
linewidth at the densities studied here).  

The simulated lineshapes were convolved with a
${\rm sinc}^2(\pi f T)$ lineshape ($T=6 \: \mu s$)
and fitted in the same manner as the experimental data
(using the ${\rm sinc}^2(\pi f T)$ lineshape convolved with a 
Lorentzian of adjustable width $\delta\nu_{dd}$).
With $n_{46p} = 5\times 10^6 \: {\rm cm}^{-3}$ 
(corresponding to $10^7 \: {\rm cm}^{-3}$ total density)
we get $\delta\nu_{dd}=18$ kHz and $\delta\nu_{dd}=63$ kHz for
cases a) and b) discussed above.  As Fig.~\ref{fg:ldens} shows,
the experimentally observed increase in $\delta\nu$ at this density 
is $55 \pm 5 \; {\rm kHz}$ -- in reasonable agreement with
the calculations.  
In making this comparison, it is
assumed that the mechanism producing the
density-dependent $\delta\nu$ observed with
no $46p_{3/2}$ atoms, is also present with the introduction
of $46p_{3/2}$ atoms, and its contribution 
is additive -- which should be reconsidered in a more precise study.
Our Rydberg density estimate is uncertain
by a factor of two, and thus an improvement in this would be desirable
for testing the limitations of this theoretical estimate
(eg. binary approximation, magnetic sub-level distributions,
constant linestrengths).

These estimates of line broadening have not accounted for motion of
the Rydberg atoms.
Consider two atoms separated by 35 $\mu$m, heading towards
each other with a relative speed of 0.4 m/s (typical of a density
of $5 \times 10^{6} \: {\rm cm}^{-3}$ and a temperature
of 300 $\mu$K).  Over the 8 $\mu$s of the experiment their
relative distance will change by only 9 \%.
The acceleration of $45d_{5/2}$ and $46p_{3/2}$ atoms towards
(or away from) one another 
-- due to the gradient of $V_{dd}$ -- should also be considered
\cite{fioretti:1999}.
The force $|\vec{F}| = |\vec{\nabla} V_{dd}| \approx 3 \mu^2/R^4$
produces a relative acceleration of 25 m/s$^2$
at a separation of 35 $\mu$m. 
Over 8 $\mu$s, this acceleration
does not change the separation appreciably -- the assumption
of interactions between stationary atoms is reasonable.
In this regard, our translationally cold Rydberg atoms are
analogous to amorphous solid-state systems \cite{macfarlane:1981}.
However, since the force due to $\hat{V}_{dd}$ has a strong
$R$ dependence, and the atoms do have some thermal motion,
there will be a (small) fraction of close
atom pairs for which movement should not be neglected,
especially at higher densities than those considered here
\cite{fioretti:1999}.

With no deliberate introduction of $46p_{3/2}$ atoms it
is observed that linewidth broadens linearly with Rydberg density
(see Fig. \ref{fg:ldens}).  
Non-resonant dipole-dipole interactions
are too weak to explain this, 
and would not show a linear density 
dependence.
We are at densities
well below the regime of spontaneous plasma formation 
\cite{killian:1999,robinson:2000} 
(we observe an ion signal that is 3 \% of the
Rydberg signal at $10^7$ cm$^{-3}$ -- too low to give
appreciable broadening due to the Stark effect from
inhomogeneous electric fields).
Trapped electron collisions \cite{dutta:2001} could broaden
the transition lines.  Again, we do not expect these to give
a linear density dependence.
The arguments of the previous paragraph rule out collisions
between cold Rydberg atoms, but 
collisions with hot Rydberg atoms from the background vapor 
may be important \cite{robinson:2000}.
At a density of $10^7 \: {\rm cm}^{-3}$
we observe significant ($\approx 10$\%) redistribution 
of the initially excited Rydberg state population into higher
angular momentum states ($l > 2$), and this redistribution scales
linearly with Rydberg density -- like the observed broadening.
The observed redistribution
rate diminishes in time following photoexcitation, as
expected for collisions with hot atoms (which diffuse from the
excitation region).
Thus we believe that either collisions with the hot Rydberg atoms,
or the products of these collisions, are responsible for the
broadening, and are currently investigating
the specific mechanism.

In summary, we have observed line-broadening in the microwave spectra
of Rydberg atoms due to resonant electric dipole-dipole interactions, 
using a combination of laser and microwave excitation
sources.  This is a general approach to the study
of cold Rydberg atom interactions.
For example, a combination of crossed optical
excitation beams (to achieve a small excitation volume) together with 
microwave transitions could allow observation of
the dipole blockade phenomena 
\cite{lukin:2001}.  

This work was supported by NSERC, CFI, and OIT.  
We thank J. Keller, J. Carter, P. Haghnegahdar
and A. Colclough for assistance.

\bibliographystyle{hieeetr}
\bibliography{references}

\end{document}